\documentclass{article}

\usepackage{amsfonts}
\usepackage{epsfig}

\def\runninghead#1#2{\pagestyle{myheadings}
\markboth{{\protect\normalsize{\quad #1}}\hfill}
{\hfill{\protect\normalsize{#2\quad}}}}
\begin{document}

\runninghead{D. Volchenkov, Ph. Blanchard}
{AN  ALGORITHM GENERATING SCALE FREE GRAPHS\hspace{1cm}}

\title{AN ALGORITHM GENERATING SCALE FREE GRAPHS }

\author{ D. Volchenkov \footnote{The Alexander von Humboldt Research Fellow at the
BiBoS Research Center, VOLCHENK@Physik.Uni-Bielefeld.DE} and  Ph. Blanchard \\
{\it  BiBoS, University Bielefeld, Postfach 100131,}\\
{\it D-33501, Bielefeld, Germany} }

\date{\today}
\maketitle

\large

\begin{abstract}
We propose a simple random process inducing various types of
random graphs and the scale free random graphs among others. The
model is of a threshold nature and differs from the preferential
attachment approach discussed in the literature before.

The degree statistics of a random graph in our model is governed
by the control parameter $\eta$ stirring the pure exponential
statistics for the degree distribution (at $\eta=0,$ when a
threshold is changed each time a new edge added to the graph) to a
power law (at $\eta=1$, when the threshold is frozen). The
exponent $\gamma$ characterizing the power law can vary in the
wide range $\gamma\in(1,\infty)$ and can be tuned in different
values $\gamma_{\mathrm{in}}$ and $\gamma_{\mathrm{out}}$ for
in-degrees and out-degrees probability distributions
independently. For the intermediate values of $\eta$, the decay
rate is mixed.

Taking different statistics for the threshold changes, one obtains
dissimilar asymptotic profiles for the degree distribution
having, in general, nothing to do with power laws at $\eta=1$, but
still uniformly exponential at $\eta=0.$

\end{abstract}
\vspace{0.5cm}

\leftline{\textbf{ PACS codes: } 02.50, 05.40-a, 89.75.-k}

\newpage

\section{Introduction}
\noindent

A wide variety of systems in biology, communication technology,
sociology and economics are best described as complex networks of
agents linked by various physical and informational links.
Clearly, the architecture of such networks is crucially important
for the vitality  and performance of these systems. It is
generally recognized that a thorough investigation of possible
mechanisms which determine the topology of complex "real-world"
networks is necessary for understanding their behavior. Numerous
studies of the Internet \cite{FFF}-\cite{GT} and World Wide Web
(WWW) \cite{AJB}-\cite{KRRT}, the web of human sexual contacts
\cite{LEASA} and movie actors collaboration networks
\cite{AJB2000}-\cite{WS}, national power-grids \cite{WS} and
phone-call networks \cite{APR}-\cite{ACL},  protein folding
\cite{ASBS,SAB} and cellular networks \cite{FW}-\cite{WF}, etc.
have revealed, among other facts, that the probability that a
node of these networks has $k$ connections follows a power law
\begin{equation}\label{powerlaw}
  P(k)\propto k^{-\gamma}
\end{equation}
over a large range of $k$, with an exponent $\gamma$ that ranges
between 1 and 3 depending on the system \cite{AB2001}. Such
networks are called scale-free \cite{BA}. Many other "real-world"
networks display an inherently uniform exponential topology that
still deviates from a Poisson distribution expected for  usual
random graphs $\mathbb{G}(N,p)$.

Recently, a model generating  a scale-free network  has been
suggested in \cite{BA}. It establishes a random graph process in
which vertices are added to the graph one at a time and connected
to a fixed number of earlier vertices, selected with probabilities
proportional to their degrees. This preferential attachment
assumption is based on the idea that a new site is more likely to
link to existing sites which are "popular" at time the site is
added. Numerical simulations  for the above process reported in
\cite{BA,BAJ} show that after many iterations the proportion of
vertices with degree $k$ indeed obeys the power law
(\ref{powerlaw}) with the exponent $\gamma=3.$ In \cite{BRST}, the
power law statistics for $P(k)$ in this model has been proven
analytically, and it  has been justified that $\gamma=3$.

Probably, a complex network formed by the citation patterns of
scientific publications where the nodes standing for published
articles and a directed edge representing a reference to a
previously published article exhibiting the  power law with the
exponent $\gamma=3$ \cite{R}, can be well fitted by the above
process as discussed in \cite{BA,BAJ,BRST}. However, in other
"real-world" networks having degree sequences matching a power
law these exponents usually deviate from $3$. For example,
$\gamma=2.3\pm 0.1$ for the movie actors collaboration network
\cite{AJB2000,ASBS}, $\gamma=2.1$ for a graph of long distance
telephone calls made during a single day \cite{APR,ACL}, etc. To
fit these deviations, one assumes that the preferential
attachment for a  real network is nonlinear \cite{JNB,N2001},
i.e. the probability that a node $i$ attaches to node $j$ can be
characterized by some function $\Pi(k)$ also following a power law
with some  index $\alpha.$

The effect of nonlinear preferential attachment  $\Pi(k)$ on the
network dynamics and topology  was studied in \cite{KRL}. They
demonstrated that the scale free network topology holds only if
$\Pi(k)$ is asymptotically linear and is destroyed otherwise.
This way the exponent of the degree distribution $P(k)$ can be
tuned to any value between 2 and $\infty$ \cite{AB2001}.

Nevertheless, the detailed studies of various "real-world"
networks still challenge to the preferential attachment approach
\cite{BA}. Some ecological networks such as food webs quantifying
the interactions between species \cite{MS}, some networks in
linguistics \cite{FCS}, a network representing an aggregation of
the WWW domains \cite{AH}, etc. exhibit a scale free topology
characterized by the indices $1<\gamma<2$. Moreover, for some
networks, the in-degree and out-degree probability distributions
are dissimilar: the WWW-topology where the nodes are the webpages
and the edges are the hyperlinks that point from one document to
another meets the power laws with $\gamma_{\mathrm{in}}=2.1$ and
$\gamma_{\mathrm{out}}=2.72$ \cite{BKMRRSTW}, the distributions
of the total number of sexual partners in a single year indicate
the power law decay with $\gamma=2.54\pm 0.2$ for females and
$\gamma=2.31\pm 0.2$ for males \cite{LEASA}. A recent study
\cite{V}  extends the results  on the citation patterns (incoming
links) of scientific publications ($\gamma_{\mathrm{in}}=3$,
\cite{R}) to the outgoing degree distribution obtaining that it
has an exponential tail (not a Poisson distribution). This
challenge stimulates a search for the new  models inducing graphs
of the scale free topology.

A random procedure which we use to generate scale free graphs
draws back to the "toy" model for a system being at a threshold of
stability reported in \cite{FVL} and to various  coherent-noise
models \cite{NS}-\cite{SN} discussed in connection to the
standard sandpile model \cite{BTW} developed in self-organized
criticality, where the statistics of avalanche sizes and
durations also take a power law form. The random process in our
model begins with no edges at time $0,$ at a chosen vertex $i,$
and adds new edges linked to $i$, one at a time; each new edge is
selected at random, uniformly among all outgoing edges not
already present. When at some time a performance of the system
occasionally exceeds a threshold value, the system looses its
stability, so that the process stops at the given vertex and then
starts again at some other vertex lacks of outgoing edges. This
way the out-degree distribution of the resulting graph equals to
the distribution of residence times below the threshold which is
closely related to the ergodic properties of the system and its
complexity, \cite{FVL}.

Although it has no a definite relation to the preferential
attachment principle \cite{BA}, probably, both algorithms are
similar: we also monitor a global state of the system (while
inspecting its stability) and introduce an auxiliary function
(which is, in some sense, analogous to $\Pi(k)$) to characterize
the  ergodic properties of the system. Nevertheless, the model we
propose in the present paper has also a distinguishing feature: we
allow for the threshold value to quench with a given probability
$\eta$, so that a natural control parameter in our model is the
relative frequency of threshold changes. Varying this frequency,
one can control the statistics of switching events (when the
system performance parameter passes the stability threshold)
tuning it from the truly exponential decay distribution (when the
threshold representing the environmental stress changes each time
a new edge added to the graph) to a power law with some exponent
$\gamma$ when the threshold is frozen.

Among clear advantages of the proposed algorithm, one can mention
its flexibility (the possible value of index $\gamma$ varies in
the wide range $(1,\infty),$ the in-degree and out-degree
statistics can be tuned independently, the variation of
frequency  $\eta$ switches $P(k)$ from an exponential decay to a
power law) as well as its relative simplicity for both analytical
and numerical studies.

The paper is organized as follows. In Sec.~2,  we describe the
model we introduce. In Sec.~3, we investigate the degree
statistics  for the random graphs generated in accordance with
the introduced random procedure. Section 6 is devoted to
conclusions.

\section{Motivation and description of the model}
\noindent

Probably, the most fascinating property of living systems is a
surprising degree of tolerance against  environmental
interventions revealing the robustness of the underlying
metabolic and genetic networks \cite{JNB,JTAOB,JMOB}. A typical
question which is usually posed in the literature in concern to
this feature is on the role of the network topology in the error
tolerance (see \cite{AB2001} and references therein). However, in
the present work, we bring an opposite question to a focus: what
is the network topology (say, of the protein-protein interactions
map) resulted from the very durable evolution process selecting
from among  species enjoying sufficiently good adaptivity? A
principle of evolutionary selection of a common large-scale
structure of biological networks  put forward in \cite{JMOB} is
closely related to this question.

The most of biological systems exhibiting a high degree of
adaptivity can be referred as systems being close to a stability
threshold, so that a slight modification of the environment would
prompt a drastic change in the system performance increasing or
reducing the stamina of biological spices. From the dynamical
systems point of view, such a behavior can be interpreted as a
forcing of bifurcation parameter through a bifurcation point
triggering the system into a new state. In our study, we do not
refer to any definite physical system displaying such a complex
behavior. Instead, following \cite{FVL}, we use a "toy" model, in
which the bifurcation parameter (the current state of the system
characterized by its net performance) and the bifurcation point
(the threshold of stability characterizing a challenge of the
environments) are considered as random independent variables. It
is supposed that the system is subjected to a bifurcation, when
its net performance takes on values above the threshold.  It is
also important that all network's changes acquired in the
previous stages of evolution of the graph are not discarded, when
the system suffers a bifurcation, but still into effect for the
forthcoming evolution.

We shall characterize the current performance of the whole system
by the real number $x\in [0,1].$ Another real number $y\in [0,1]$
plays the role of the stability threshold. The network  is
supposed to be stable until $x<y$ and is condemned otherwise. We
consider $x$ as a random variable distributed with respect to
some given probability distribution function (pdf) $f(u):[0,1]\to
\mathbb{R}^+$. In a deterministic version of the model, $f$ would
be an ergodic invariant density function on the phase space which
has an essential influence on the system dynamics close to the
stability threshold value.

The value of threshold $y$ also fluctuates reflecting the changes
of environments and is also treated as a random variable
distributed over the interval $[0,1]$ with respect to some other
pdf $g(u):[0,1]\to\mathbb{R}^+$. In general, $f$ and $g$ are two
arbitrary positive integrable functions satisfying the
normalization condition
\begin{equation}\label{norm}
  \int^1_0 du {\ } f(u)=\int^1_0 du {\ }g(u)=1.
\end{equation}
We define the  random  procedure on the fixed set of  $N$ vertices
describing the evolution of network as a random graph evolving in
time. The process begins with no edges at time $0,$ at a chosen
vertex $i$. Given a fixed real number $\eta \in[0,1]$, we define a
discrete time random process in the following way. At time $t=0,$
the variable $x$ is chosen with respect to pdf $f$, and $y$ is
chosen with pdf $g$. If $x<y$, we add an edge selected at random,
uniformly among all edges outgoing from $i$, the process continues
and goes to time $t=1$. At time $t\geq 1,$ one of  the following
events happens:

\textbf{i}) with probability $\eta$, the random variable $x$ is chosen
with pdf $f$ but the threshold $y$ keeps the value it had at time
$t-1$. Otherwise,

\textbf{ii}) with probability $1-\eta,$ the random variable $x$ is chosen
with pdf $f,$ and  the threshold $y$ is chosen with pdf $g$.

If $x\geq y,$ the process at the given  vertex ends and then
resumes at some other vertex chosen randomly from the vertices
having no outgoing edges yet; if $x<y,$ the process continues at
the same vertex and goes to time $t+1.$

Eventually, when at some time $T$ limiting the duration of the
"monotonous" evolution of the graph, the process moves to some
other vertex, and $i$ has exactly $k=T$ outgoing edges.

For example, in Fig.~1  we have presented the results of numerical
simulation for the above algorithm for a network consisting of
$2\times 10^3$ vertices, for $f=g=1$ and $\eta=1$. It displays the
proportion of vertices having degree $k$ versus $k$ in the
log-log scale. Circles stay for the out-degree $k_{\mathrm{out}}$
and diamonds are for the in-degree $k_{\mathrm{in}}.$ One can see
that the out-degree probability distribution is fitted well by a
power law, while the in-degree probability distribution forms a
Gaussian profile centered at the most probable number of incoming
links $k^*_{\mathrm{in}}=11.$

The same algorithm can be extended readily to moderate the
incoming edges also. Let us consider three random variables
$x,y,$ and $z$ that are the real numbers distributed in
accordance to the distributions $f,$ $g$, and $v$ within the unit
interval $[0,1].$ We assume that $x$ represents the current
performance of the system, $y$ and $z$ are the thresholds for
emission and absorption of edges respectively.

The generalized process begins on the set of $N$ vertices with no
edges at time $0,$ at a chosen vertex $i$. Given two fixed numbers
$\eta\in[0,1]$ and $\nu\in [0,1]$, the variable $x$ is chosen
with respect to pdf $f$, $y$ is chosen with pdf $g$, and $z$ is
chosen with pdf $v$, we draw $e_{ij}$ edge outgoing from $i$
vertex and entering $j$ vertex if $x<y$ and $x<z$ and continue
the process to time $t=1.$ Otherwise, if $x\geq y$ ($x\geq z$),
the process moves to other vertices having no outgoing (incoming)
links yet.

At time $t\geq 1,$ one of the three events happens:

\textbf{i}) with probability $\eta$, the random variable $x$ is chosen
with pdf $f$ but the thresholds $y$ and $z$ keep their values
they had at time $t-1$.

\textbf{ii}) with probability $1-\eta,$ the random variable $x$ is chosen with
pdf $f,$ and the thresholds $y$ and $z$ are chosen with pdf $g$
and $v$ respectively.

\textbf{iii}) with probability $\nu,$ the random variable $x$ is chosen with pdf
$f$, and the threshold $z$ is chosen with pdf $v$  but the
threshold $y$ keeps the value it had at time $t-1$.

If $x\geq y$, the process stops at $i$ vertex and then starts at
some other vertex having no outgoing edges yet. If $x\geq z,$ the
accepting vertex $j$ is blocked and does not admit any more
incoming link (provided it has any).  If $x<y$ and $x<z$, the
process continues at the same vertex $i$ and goes to time $t+1.$

Figure 2 displays an example of a graph generated in accordance
this algorithm on the set of 100 vertices for the case of uniform
densities $f=g=v=1$ for $\eta=\nu=1.$ In Fig.~3, we have presented
the distribution $P(k)$ vs. $k$ in the log-log scale over $10^5$
vertices. Here, the circles stay for outgoing degrees, and
diamonds are for incoming degrees. For $k\gg 1,$ both profiles
enjoy a power law decay with
$\gamma_{\mathrm{in}}=\gamma_{\mathrm{out}}=2.$

In the forthcoming section, we demonstrate that for some
probability densities $f,g$, and $v$ the resulting probability
distribution $P(k)$ meets asymptotically a power law
(\ref{powerlaw}) as $k\gg 1$ where the index $\gamma$ can vary in
the wide range $\gamma\in (1,\infty)$. Moreover, by means of
adjusting probability densities $f,g,$ and $v$ from a proper
class of functions, one can tune $\gamma_{\mathrm{in}}$ and
$\gamma_{\mathrm{out}}$ to different values.

If $\nu=0$, both thresholds $y$ and $z$ have synchronized
dynamics, and sliding the value of $\eta$ form 0 to 1, one can
tune the statistics of out-degrees and in-degrees simultaneously
out from the pure exponential decay to the power laws provided
$f,$ $g$, and $v$ belong to the proper class of functions. When
$\nu=1$, the threshold $z$ changes each time but the threshold
$y$ is frozen. In this case, $P(k_{\mathrm{in}})$ decays
exponentially for any choice of $v$, but $P(k_{\mathrm{out}})$
still meets a power law for $k_{\mathrm{out}}\gg 1$.

\section{Probability degree distribution $P(k)$}
\noindent

Below, we compute the probability $P(k)$ that a given vertex has
precisely $k$ outgoing edges as a result of the random procedure
defined in the previous section. For sake of clarity, we
investigate the statistics of switching events quenching the
monotonous evolution of the graph in the simplest case when only
one threshold ($y$) is present. All computations can be extended
readily to the general case when both thresholds are present.

Directly from the definitions of Sec.~2, one can show that
$$
  P(0)=\int^1_0 dy {\ }g(y)\int^1_y dx {\ }f(x).
$$
In $t\geq 1$ time steps, the system can either "survive" (S) or
"die" (D). Both issues can take place either in the "correlated"
way (with probability $\eta$, see \textbf{i}) (we denote these
scenarios as $S_c$ and $D_c$ consequently), or in the
"uncorrelated" way (with probability $1-\eta$, see \textbf{ii})
($S_u$ and $D_u$, consequently). For example, for $k=1$, we have
$$
  P(1) =  P[SD_c]+P[SD_u]=\eta B(1)+(1-\eta)A^2(1)
$$
where we have defined, for $n=0, 1, 2, \ldots,$
\begin{equation}\label{A}
  A(n)=\int_0^1 dy {\ } g(y)\left[\int^y_0 dx{\ }f(x) \right]^n
\end{equation}
and
\begin{equation}\label{B}
B(n)=\int_0^1 dy {\ } g(y)\left[\int^y_0 dx{\ }f(x) \right]^n
\int^1_y du {\ } f(u)= A(n)-A(n+1).
\end{equation}
Similarly,
$$
  P(2)=P[SS_cD_c]+P[SS_cD_u]+P[SS_uD_c]+P[SS_uD_u],
$$
where
$$
P[SS_cD_c]= \eta^2B(2), \quad P[SS_cD_u]=\eta(1-\eta)A(2)B(0),
$$
$$
P[SS_uD_c]=\eta(1-\eta)A(1)B(1), \quad P[SS_uD_u]=(1-\eta)^2
A^2(1)B(0).
$$
The general formula for $k\geq 3$ is tedious and can be found in
\cite{FVL}.

Computations can be essentially simplified with the use of
generating functions. Let us define the generating function for
$P(k)$ by
\begin{equation}\label{inverse}
 \hat{P}(s)=\sum_{k=0}^{\infty}s^kP(k), \quad
 P(k)=\left.\frac 1{k!}\frac{d^k \hat{P}(s)}{ds^k}\right|_{s=0},
\end{equation}
and introduce the new auxiliary functions,
$$
p(l)=\eta^l A(l+1), \quad \mathrm{for } {\ } l\geq 1, \quad
p(0)=0,
$$
$$
q(l)=(1-\eta)^l A^{l-1}(1), \quad \mathrm{for } {\ } l\geq 1,
\quad q(0)=0,
$$
$$
r(l)=\eta^l\left[\eta B(l+1)+ (1-\eta) A(l+1) B(0)\right], \quad
\mathrm{for } {\ } l\geq 1, \quad r(0)=0,
$$
\begin{equation}\label{pqr}
  \rho=\eta B(1)+ (1-\eta)A(1)B(0).
\end{equation}
Then, the use of convolutions property \cite{conv} yields
\begin{equation}\label{PS}
 \hat{P}(s)= B(0)+\rho s+ \frac
 {s}{1-\hat{p}(s)\hat{q}(s)}\left[
\hat{r}(s)+\rho\hat{p}(s)\hat{q}(s)+\rho A(1)\hat{q}(s)+A(1)\hat{q}(s)\hat{r}(s)
 \right],
\end{equation}
where $\hat{p}(s),\hat{q}(s),\hat{r}(s)$ are the generating
functions of $p(l), q(l), r(l),$ respectively.

In the marginal cases of $\eta=0$ and $\eta=1$,  the probability
$P(k)$ can be readily calculated.

\subsection{The uncorrelated case, $\eta=0$}
\noindent

This case corresponds to the external stress being sensitive to
any modification of a network and changing its figure as soon as a
new edge appears in the graph. For $\eta=0,$ one has
$\hat{p}(s)=\hat{r}(s)=0,$ and $\hat{q}(s)=s(1-s A(1))^{-1},$ ${\
} \rho=A(1)B(0)$, then equation (\ref{PS}) gives
\begin{equation}\label{eta0}
\hat{P}_{\eta =0}(s)= \frac{B(0)}{1-sA(1)}.
\end{equation}
Applying the inverse formula in (\ref{inverse}) to equation
(\ref{eta0}), one arrives at
$$
  P_{\eta=0}(k)=A^k(1)B(0) =\left[
\int ^1_0 dy {\ } g(y)\int^y_0 dx{\ } f(x)
  \right]^k\int^1_0 dy {\ }g(y)\int^1_y dx {\ } f(x).
$$
Therefore, in the "uncorrelated" case ($\eta=0$), for \textit{any
choice} of the pdf $f(u)$ and $g(u),$ the probability $P(k)$
decays exponentially.

\subsection{The correlated case, $\eta=1$}
\noindent

Another marginal case corresponds to the external stress
remaining unchanged during all time of  network evolution. For
$\eta=1$, one has $\hat{q}(s)=0,$ ${\ }
\rho =B(1),$ and
$$
s\hat{r}(s)=\sum^\infty_{n=1}s^{n+1}B(n+1)=\sum^\infty_{n=2}s^nB(n),
$$
so that formula (\ref{PS}) yields
\begin{equation}\label{eta1}
  \hat{P}_{\eta=1}(s)=\hat{B}(s).
\end{equation}
Consequently,
\begin{equation}\label{Peta1}
  P_{\eta=1}(k)=B(k)=\int_0^1 dy {\ }g(y)\left[\int _0^y dx {\
  }f(x)\right]^k
  \int_y^1 du \; f(u).
\end{equation}
In the "correlated" case, $\eta=1$, many different types of
behaviour are possible, depending on the form of the pdf $f(u)$
and $g(u)$. We will examine an important class of $f(u)$ and
$g(u)$, for which $P_{\eta=1}(k)$ can be explicitly computed from
equation (\ref{Peta1}). We will take
\begin{equation}\label{ss}
  \begin{array}{lc}
   f(u) = (1+\alpha) u^{\alpha}, & \alpha > -1, \\
  g(u) = (1+\beta) (1-u)^{\beta},& \beta > -1.
  \end{array}
\end{equation}
Equation (\ref{Peta1}) gives in this case:
$$
P_{\eta=1}(k) = \frac{\Gamma(2+\beta) \; \Gamma(1 + k (1+\alpha))}
    {\Gamma(2 + \beta + k (1+\alpha))}
- \frac{\Gamma(2+\beta) \; \Gamma(1 + (k+1) (1+\alpha))}
    {\Gamma(2 + \beta + (k+1) (1+\alpha))} \;.
$$
Using Stirling's approximation, we get for large $k>>1$:
$$
P_{\eta=1}(k) = \frac{(1+\beta)\; \Gamma(2+\beta)\;
(1+\alpha)^{-1-\beta}}
    {k^{2+\beta}}\; \left( 1+0\left(\frac{1}{k}\right) \right)\;.
$$
For different values of $\beta$, the exponent of the threshold
distribution, we get all possible power law decays of
$P_{\eta=1}(k)$. Notice that the exponent $(-2-\beta)$
characterizing the decay of $P_{\eta=1}(k)$ is independent of the
distribution $f(u)$ of the state variable.

Therefore, for $\eta=1$, one has to provide just one power law
function (which defines the statistics of the threshold) to
achieve a power law for $P(k)$ characterized by the exponent
$\gamma\in(1,\infty).$

\subsection{The case of uniform densities $f=g=1$}
\noindent

For the case of uniform densities $f=g=1,$ and after some tedious
but trivial computation, we get from equation (\ref{PS}):
\begin{equation}\label{Psfg1}
  \hat{P}(s)=\frac 1{1+(1-\eta)\mu(s)}\left[
  \frac{1+\mu(s)}{s}-\eta\mu(s)\right]\;,
\end{equation}
where $\mu(s)$ is defined by
$$
\mu(s)=\frac{\ln(1-\eta s)}{\eta s}\;.
$$
The asymptotic behavior of $P(k)$ for $k$ large is determined by
the singularity of the generating function $\hat{P}(s)$ that is
closest to the origin.

For $\eta=0,$ the generating function has a simple pole
$\hat{P}(s)=(2-s)^{-1},$ and therefore $P(k)$ decays
exponentially, which agrees with the general conclusion of
Sec.~3.1.

For intermediate values $1>\eta>0$, the generating function
$\hat{P}(s)$ has two singularities. One pole, $s=s_0,$ corresponds
to the vanishing denominator $1+(1-\eta)\mu(s),$ where $s_0$ is
the unique nontrivial solution of the equation
\begin{equation}\label{s0}
  -\ln(1-\eta s)=s\frac{\eta}{1-\eta}\;.
\end{equation}
Another singularity, $s=s_1,$ corresponds to the vanishing
argument of the logarithm, $s_1=\eta^{-1},$ such that
$s_1>s_0>1.$ The dominant singularity of $\hat{P}(s)$ is of polar
type, and the corresponding decay of $P(k)$ is asymptotically
exponential for the degrees $k$ larger than $k_c \sim
\ln(s_0)^{-1}$, with a rate $\ln(s_0)$ that vanishes like
$1-\eta$ as $\eta$ tends to 1.

When $\eta$ tends to one, two singularities $s_0$ and $s_1$ merge.
More precisely, we have
\begin{equation}\label{Pseta1}
\hat{P}_{\eta=1}(s)=\frac{s+(1-s)\ln(1-s)}{s^2}.
\end{equation}
The corresponding dominant term in (\ref{Pseta1}) is of order
$\mathrm{O}(k^{-2})$ \cite{Flajolet}. This obviously agrees with
the exact result one can get from equation (\ref{Peta1}), with
$f(x)=g(x)=1$:
\begin{equation}\label{PTeta1}
P_{\eta=1}(k)=\frac {1}{(k+1)(k+2)}\;.
\end{equation}
In the case of uniform densities, it is possible to get an
expression of $P(k)$ for all $k$, and for any value of $\eta$, by
applying the inversion formula (\ref{inverse}) to equation
(\ref{Pseta1}) and use of the formulae \cite{formulae}. Combining
all this, we get
\begin{equation} \label{etavary1}
P(k) = \frac{\eta^k}{(k+1)(k+2)} +
    \sum_{l=1}^{k} \frac{\eta^k}{(k-l+1)(k-l+2)\,l}\,\sum_{m=1}^{l}
    \left(\frac{1-\eta}{\eta}\right)^m  c(m,l)\;,
\end{equation}
where $c(m,l)$ is defined by
\begin{eqnarray*}
c(m,l) = m! \, \sum_{\begin{array}{c}
    \scriptstyle l_1 + l_2 + \cdots + l_{m} \,= \,l\\
        \scriptstyle l_i \ge 1
    \end{array}}  \hspace{10cm}\\
\frac{l_1 \, l_2 \, \cdots \, l_{m-1} \, l_m}
    {(l_1+1) \, (l-l_1) \, (l_2+1) \, (l-l_1-l_2) \, \cdots \,
    (l_{m-1}+1) \, (l-l_1-l_2-\cdots-l_{m-1}) \, (l_m+1)} \;.
\end{eqnarray*}
When $\eta\neq 0$, there is an alternative way of writing the
previous expression:
\begin{equation} \label{etavary2}
P(k) = \frac{\eta^k}{(k+1)(k+2)} +
    \sum_{l=1}^{k} \frac{\eta^{k+1}}{(k-l+1)(k-l+2)}\,
    \sum_{m=1}^{\infty} (1-\eta)^m  \, b(m,l)\;,
\end{equation}
where $b(m,l)$ is defined by
$$
b(m,l) = \sum_{\begin{array}{c}
    \scriptstyle i_1 + i_2 + \cdots + i_{l} \,= \,l\\
        \scriptstyle i_j \ge 0
    \end{array}}
\frac{1}{(i_1+1) \, (i_2+1) \, \cdots \, (i_{m-1}+1) \, (i_m+1)} \;.
$$
In Fig.~4, we have plotted these asymptotes in the log-linear
scale  for the consequent values $\eta=0,$ ${\ }\eta = 0.5$, ${\
}\eta = 0.7$, ${\ }\eta = 0.9$, ${\ }\eta=1$ (bottom to top)
together with the curves representing analytical results
(\ref{PTeta1}-\ref{etavary1}).

Notice that for $\eta \neq 1$, we only plot distributions $P(k)$
up to relatively small degrees ($k=16$, $k=25$) correspondent to
short quiescent times in the system, since these times are already
bigger then the crossover value $k_c(\eta)\sim 1/\ln(s_0)$ to the
exponential decay $\exp{(-\,\ln(s_0)\,k)}$ (where $s_0$ is defined
by the Eq. (\ref{s0})). For much longer times, very few survivals
are observed, and the statistics gets bad. Of course, $k_c(\eta)$
grows as the parameter $\eta$ tends to 1, so that we have good
statistics for larger $k$.

\section{Discussion and Conclusion}
\noindent

In this paper, we have presented a simple random procedure
inducing various types of random graphs and the scale free random
graphs among others. The model we use essentially differs from the
preferential attachment approach \cite{BA} discussed in the
literature before. The proposed approach is simple and more
flexible both for numerical simulations and analytical studies.
The model draws back to the original "toy" model  for a system
being at a threshold of stability \cite{FVL}.

In the proposed model, the degree statistics of a random graph
depends from the value of the control parameter $\eta$ stirring
the pure exponential statistics of degrees (at $\eta=0,$ when the
threshold value is sensitive to each new edge risen in the graph)
to a power law (at $\eta=1$, when the threshold value is frozen).
The exponent characterizing the power law depends as $\gamma
=2+\beta$ from the power law probability density distribution of
the threshold $g(u)\propto (1-u)^{\beta}, $ ${\ } \beta>-1,$ and
can vary in the wide range $\gamma\in(1,\infty).$ For the
intermediate values of $\eta$, the decay rate is mixed. In
particular, we showed that, even if the asymptotic behavior is
exponential, when $\eta$ is close to 1 this only holds for very
large degrees $k$ of order of $(1-\eta)^{-1}$. The model can be
readily generalized to various particular types of networks. The
crucial advantage of the model is that, for most physically
meaningful distributions, it can be solved analytically. In other
cases, a numerical solution can be readily found.

In the present paper, we have limited the consideration to the
special cases of uniform and power law distributions of
thresholds. Indeed, one can take other (normalized) distributions
for $f,g,$ and $v$ in this model and obtain different decaying
asymptotic profiles for $P(k)$ either for synchronous $\nu=0$ or
asynchronous $\nu=1$ dynamics of thresholds,  they have, in
general, nothing to do with power laws at $\eta=1$, but still
uniformly exponential at $\eta=0.$

\section{Acknowledgements}
\label{07}
\noindent

One of the authors (D.V.)  benefits from a scholarship of the
Alexander von Humboldt Foundation (Germany) that he gratefully
acknowledges.

\newpage

\begin{figure}[hp]
 \noindent
 \begin{minipage}[b]{.36\linewidth}
 \begin{center}
 \epsfig{file=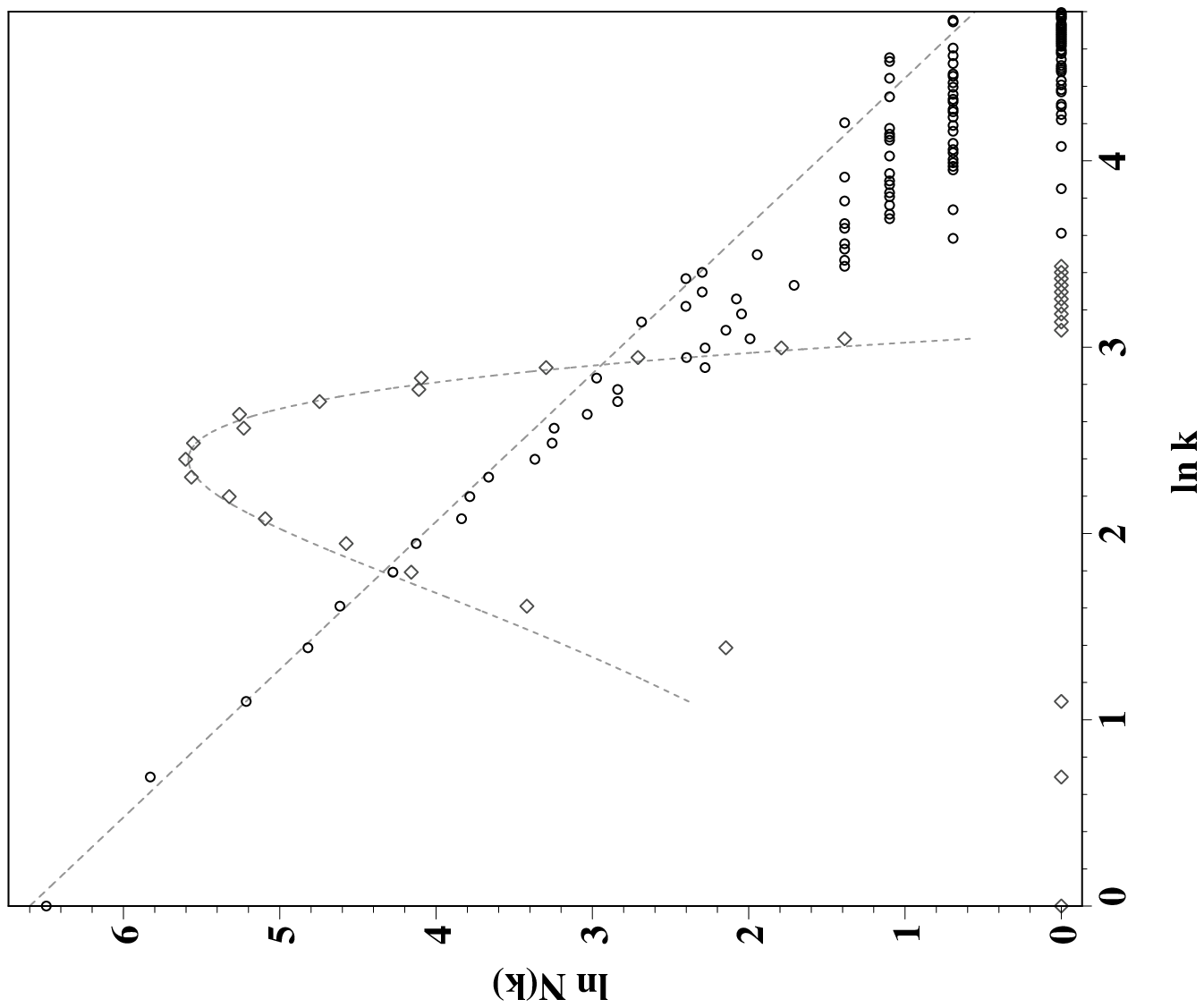, angle= -90}
 \end{center}
\end{minipage}
\label{1}
\caption{The proportion of vertices of degree $k$ versus $k$ in
the log-log scale ($N=2000,$ ${\ }f=g=1,$ ${\ } \eta=1$).
Out-degree statistics (circles) satisfies a power law, while the
in-degree probability distribution is Gaussian with the most
probable number of incoming links $k^*_{\mathrm{in}}=11.$}
\end{figure}

\begin{figure}[hp]
 \noindent
 \begin{minipage}[b]{.36\linewidth}
 \begin{center}
 \epsfig{file=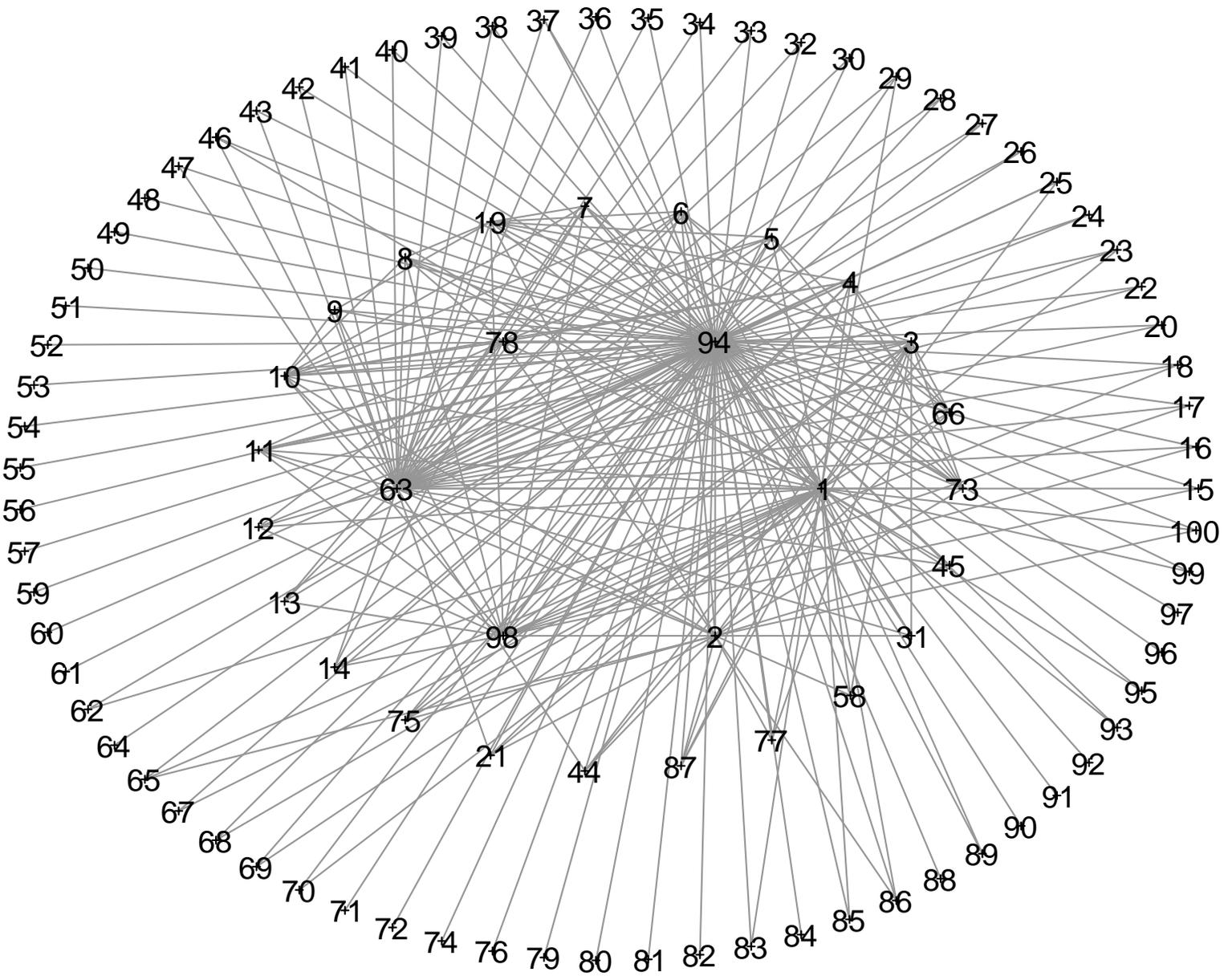, angle= 0}
 \end{center}
\end{minipage}
\label{2}
\caption{An example of a graph generated with the
algorithm of Sec.~2 on the set of 100 vertices for the case of
uniform densities $f=g=v=1,$ for $\eta=1.$}
\end{figure}

\begin{figure}[hp]
 \noindent
 \begin{minipage}[b]{.36\linewidth}
 \begin{center}
 \epsfig{file=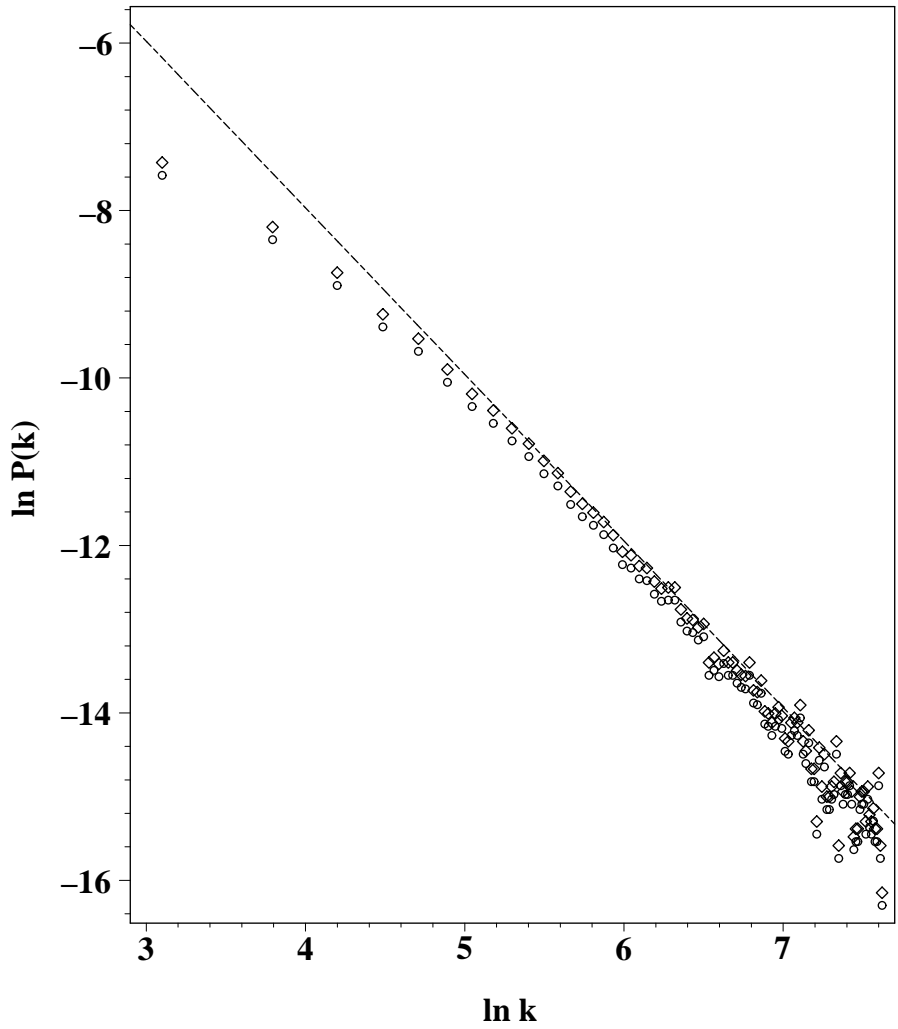, angle= -90}
 \end{center}
\end{minipage}
\label{3}
\caption{The distribution $P(k)$ vs. $k$ in the log-log scale generated
on $10^5$ vertices, $f=g=v=1,$ ${\ }\eta=1$. Here, the circles
stay for outgoing degrees and diamonds are for incoming degrees.
Both profiles enjoy a power law decay with $\gamma=2.$}
\end{figure}

\begin{figure}[hp]
 \noindent
 \begin{minipage}[b]{.36\linewidth}
 \begin{center}
 \epsfig{file=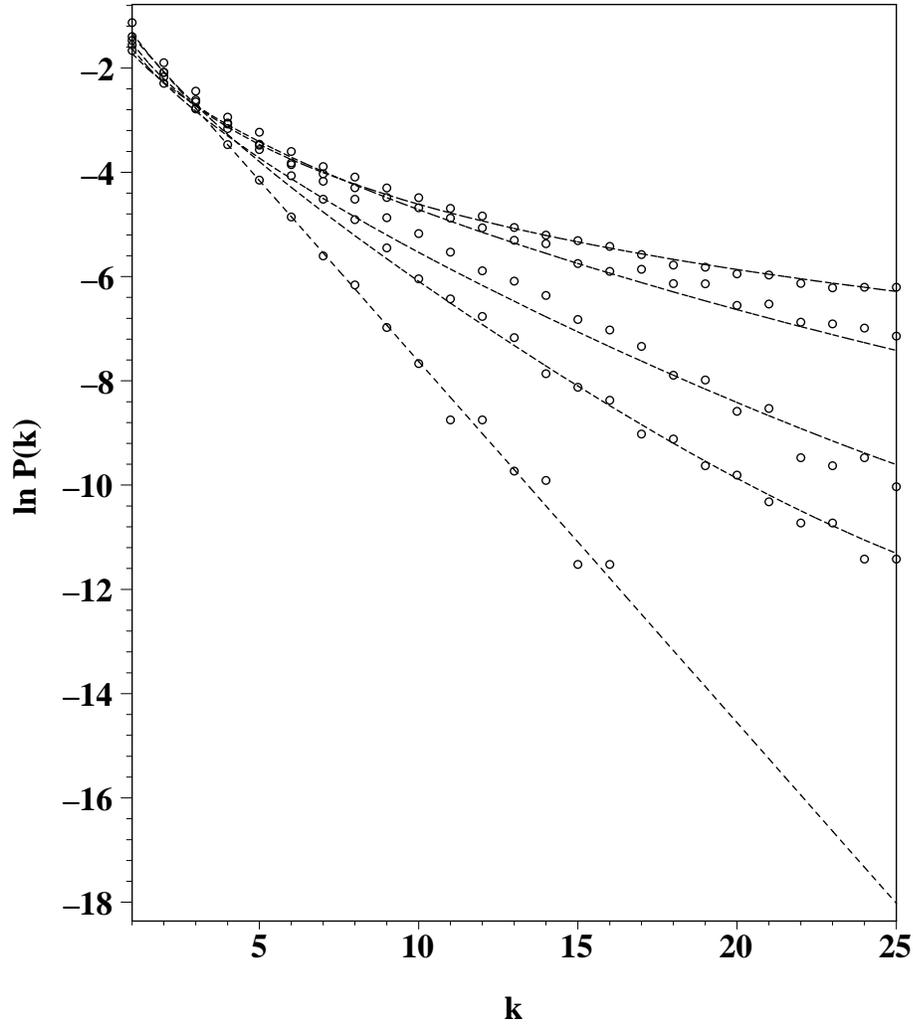, angle= -90}
 \end{center}
\end{minipage}
\label{4}
\caption{The distribution $P(k)$ in
log-linear scale for different values of the parameter $\eta$,
${\ }\eta=0.,$ ${\ }\eta=0.5,$ ${\ }\eta=0.7,$ ${\ }\eta=0.9$ and
$\eta=1.0$ (bottom to top).  Straight lines corresponds to the
pure exponential decay ($\sim 2^{-k}$) observed in the case
$\eta=0$, the results obtained from (\ref{etavary1}) for the
intermediate values of $\eta$, and to the result (\ref{PTeta1})
for $\eta=1.0$ (bottom to top).}
\end{figure}


\begin{thebibliography}{000}
\bibitem{formulae}

We use the following facts for the computations:
$$
\left.\frac{d^n \gamma(s)}{ds^n}\right|_{s=0} = - \frac{n!}{n+1}\,\eta^n \;,
$$
$$
\left.\frac{d^n}{ds^n} \frac{1+\gamma(s)}{s} \right|_{s=0} =
- \frac{n!}{n+2}\,\eta^{n+1} \;,
$$
$$
\frac{d^n}{ds^n} f(s)g(s) = \sum_{k=0}^{n}
\left( \begin{array}{c} n \\ k \end{array} \right)
    \frac{d^k f(s)}{ds^k} \, \frac{d^{n-k} g(s)}{ds^{n-k}} \;,
$$
$$
\frac{d^n}{ds^n} \frac{1}{1-f(s)} = \frac{f^{(n)}(s)}{[1-f(s)]^2} +
    (n-1)! \sum_{m=2}^{n} \frac{m!}{[1-f(s)]^{m+1}}
    \sum_{\begin{array}{c}
    \scriptstyle l_1 + l_2 + \cdots + l_{m} \,= \,n\\
        \scriptstyle l_i \ge 1
    \end{array}}
$$
$$
    \frac{f^{(l_1)}(s) \, {f^{(l_2)}(s)} \, \cdots \, {f^{(l_{m-1})}(s) \, f^{(l_m)}(s)}}
    {(l_1-1)! \, (n-l_1) \, (l_2-1)! \, (n-l_1-l_2) \, \cdots \, (l_{m-1}-1)! \,
    (n-l_1-l_2-\cdots-l_{m-1}) \, (l_m-1)!}\;,
$$
where the latter equation is valid for all $n\ge 1$.
\end{thebibliography}
\end{document}